\documentclass[iop]{emulateapj-rtx4}

\newcommand{\logr}{log(\ensuremath{R'_{\mbox{\scriptsize HK}}})}
\newcommand{\Mj}{M$_\textrm{\scriptsize J}$}
\newcommand{\Teff}{T$_\textrm{\scriptsize eff}$}
\newcommand{\Teq}{T$_\textrm{\scriptsize eq}$}

\usepackage{natbib}
\usepackage{upgreek}
\usepackage{color}          
\usepackage{wasysym}                                                                      

\begin{document}
\bibliographystyle{plainnat}

\title{
$HST$ hot-Jupiter transmission spectral survey: \\Clear skies for cool Saturn WASP-39\lowercase{b}
}
~\\
\author{
Patrick D. Fischer\altaffilmark{1}, 
Heather A. Knutson\altaffilmark{1}, 
David K. Sing\altaffilmark{2}, 
Gregory W. Henry\altaffilmark{3},
Michael W. Williamson\altaffilmark{3},
Jonathan J. Fortney\altaffilmark{4}, 
Adam S. Burrows\altaffilmark{5},
Tiffany Kataria\altaffilmark{2},
Nikolay Nikolov\altaffilmark{2},
Adam P. Showman\altaffilmark{6},
Gilda E. Ballester\altaffilmark{6},
Jean-Michel D\'esert\altaffilmark{7}, 
Suzanne Aigrain\altaffilmark{8},
Drake Deming\altaffilmark{9},
Alain Lecavelier des Etangs\altaffilmark{10},
Alfred Vidal-Madjar\altaffilmark{10}
}

\affil{\altaffilmark{1}Division of Geological and Planetary Sciences, California Institute of Technology, Pasadena, CA 91125, USA}
\affil{\altaffilmark{2}Astrophysics Group, School of Physics, University of Exeter, Stocker Road, Exeter, EX4 4QL, UK}
\affil{\altaffilmark{3}Center of Excellence in Information Systems, Tennessee State University, Nashville, TN 37209, USA}
\affil{\altaffilmark{4}Department of Astronomy and Astrophysics, University of California Santa Cruz, CA 95064, USA}
\affil{\altaffilmark{5}Department of Astrophysical Sciences, Peyton Hall, Princeton University, Princeton, NJ 08544, USA}
\affil{\altaffilmark{6}Lunar and Planetary Laboratory, University of Arizona, Tucson, AZ 85721, USA}
\affil{\altaffilmark{7}CASA, Department of Astrophysical and Planetary Sciences, University of Colorado, 389-UCB, Boulder, CO 80309, USA}
\affil{\altaffilmark{8}Department of Physics, University of Oxford, Denys Wilkinson Building, Keble Road, Oxford OX1 3RH, UK}
\affil{\altaffilmark{9}Department of Astronomy, University of Maryland, College Park, MD 20742, USA}
\affil{\altaffilmark{10}CNRS, Institut dAstrophysique de Paris, UMR 7095, 98bis boulevard Arago, F-75014 Paris, France}

\begin{abstract}

We present $HST$ STIS optical transmission spectroscopy of the cool Saturn-mass exoplanet WASP-39b from 0.29-1.025 $\upmu$m, along with complementary transit observations from $Spitzer$ IRAC at 3.6 and 4.5 $\upmu$m. The low density and large atmospheric pressure scale height of WASP-39b make it particularly amenable to atmospheric characterization using this technique. We detect a Rayleigh scattering slope as well as sodium and potassium absorption features; this is the first exoplanet in which both alkali features are clearly detected with the extended wings predicted by cloud-free atmosphere models. The full transmission spectrum is well matched by a clear, H$_2$-dominated atmosphere or one containing a weak contribution from haze, in good agreement with the preliminary reduction of these data presented in \citet{sing16}. WASP-39b is predicted to have a pressure-temperature profile comparable to that of HD~189733b and WASP-6b, making it one of the coolest transiting gas giants observed in our $HST$ STIS survey. Despite this similarity, WASP-39b appears to be largely cloud-free while the transmission spectra of HD~189733b and WASP-6b both indicate the presence of high altitude clouds or hazes. These observations further emphasize the surprising diversity of cloudy and cloud-free gas giant planets in short-period orbits and the corresponding challenges associated with developing predictive cloud models for these atmospheres.

\end{abstract}


\section{Introduction}

Over the past twenty years, ongoing radial velocity and transit surveys have detected more than 200 short-period gas giant planets transiting bright, nearby stars. This population of planets, which are often referred to as hot Jupiters, has provided an unprecedented opportunity to study the properties of hydrogen-dominated atmospheres at temperatures more akin to cool stars than solar system gas giants \citep{sea10, hen15}. Despite more than a decade of study, many aspects of hot Jupiter atmospheres remain poorly understood, as demonstrated by the detection of clouds and hazes in a subset of these atmospheres. 

We can determine the presence or absence of clouds in hot Jupiter atmospheres via several complementary techniques. Observations of the secondary eclipse, when the planet passes behind the star, can be used to constrain the planet's visible-light albedo and detect the signatures of reflective cloud layers in the upper atmosphere \citep[e.g.][]{hen13, eva13, she14, ang14}. By observing changes in the planet's albedo as a function of orbital phase, we can map the relative locations of these reflective cloud layers \citep{demo13, shp15}. These observations indicate that there is a range of hot Jupiter albedos, consisting of planets with relatively low ($<$0.1) to somewhat higher (0.3-0.4) albedos in the optical Kepler bandpass. 

Although a high albedo can indicate the presence of clouds, some planets may have relatively tenuous and/or low-albedo cloud layers. By measuring the wavelength-dependent transit depth or transmission spectrum of these planets as they pass in front of their host stars, we can detect the presence of trace clouds and hazes located near the planet's day-night terminator. During the transit light from the star travels along a slant optical path through the upper part of the planet's atmosphere, where even relatively small amounts of haze or cloud particles can result in a significant scattering opacity \citep[e.g.][]{for03, for05, pon08}. For cloud-free hot Jupiters, sodium and potassium are predicted to produce strong absorption at optical wavelengths, while water is the strongest absorber in the near-infrared \citep[e.g.][]{sea00, hub01, sud03}. However in many of the systems observed, the predicted absorption features from Na, K, and water are either attenuated or entirely absent and the transmission spectrum displays a strong slope across the optical wavelengths \citep[e.g.][]{lec08, pon08, sing11, hui12, pon13, dem13, line13, man13, knu14, mcc14, nik14, nik15, sing13, sing15, sing16}.

It has been suggested that the high-altitude clouds or hazes responsible for these attenuated absorption features may be produced via photochemistry at the top of the atmosphere \citep{zah09} or by condensation within the atmosphere \citep[e.g.][]{for05, lec08,mor13, wak15}. However, the mechanisms that drive cloud formation in hot Jupiter atmospheres are not fully understood, and may involve horizontal or vertical transport of materials from the planet's night side and deeper atmosphere in addition to the atmospheric metallicity, surface gravity, and local pressure and temperature. For photochemically produced hazes formation rates might additionally depend on the stellar spectral type and activity level, which controls the incident UV flux at the top of the atmosphere \citep{zah09, knu10}. As discussed in \citet{sing16}, the current set of transmission spectroscopy observations appear to be poorly matched by the predictions of simple forward models, suggesting that our knowledge of the factors that contribute to or suppress cloud formation in these atmospheres is incomplete.

In this study we present observations from an ongoing survey of optical transmission spectra of hot Jupiters obtained with the Space Telescope Imaging Spectrograph (STIS) instrument on the $Hubble~Space~Telescope$ ($HST$). The goal of this survey is to build up a large sample of hot Jupiters with well-characterized transmission spectra in order to develop a better empirical understanding of the relevant factors that determine the presence or absence of clouds in these atmospheres \citep{hui12, wak13, nik14, nik15, sing13, sing15,sing16}. Here we examine in detail the transmission spectrum of the hot Jupiter WASP-39b originally presented in \citet{sing16}. WASP-39b \citep{fae11} is relatively cool with an equilibrium temperature of 1120~K assuming zero albedo and efficient redistribution of energy to the night side. Recently published secondary eclipse measurements at 3.6 and 4.5 micron are in good agreement with these assumptions, although there is an intrinsic degeneracy between the assumed albedo and the atmospheric circulation efficiency when interpreting dayside emission spectra \citep{kam15}. WASP-39b is approximately Saturn-mass (0.28~\Mj) with an inflated radius of 1.27~R$_\textrm{\scriptsize J}$, making it one of the lowest density gas giant planets currently known (0.14~$\rho_\textrm{\scriptsize J}$), and particularly favorable for atmospheric characterization via transmission spectroscopy. It orbits at 0.049 AU around a relatively quiet G8 star with an effective temperature of 5400~K and [Fe/H]=-0.12$\pm$0.10. In the following sections we present STIS transit observations of this planet spanning wavelengths between 290-1025~nm as well as 3.6 and 4.5~$\upmu$m photometry obtained with the $Spitzer~Space~Telescope$, comprising a high signal-to-noise near-UV to infrared transmission spectrum. 

\section{Observations}

\subsection{$HST$ STIS}

Observations of WASP-39b in transmission were obtained with $HST$ Space Telescope Imaging Spectrograph (STIS) as part of $HST$ program GO-12473 (P.I. Sing). Two transits of WASP-39b were observed in the G430L grating (290-570~nm) on UT 2013 Feb 8 and 12, and one transit with the G750L grating (550-1020~nm) on UT 2013 Mar 17. These observations span 11 orbital periods of WASP-39b, approximately 45 Earth days. Each observation consists of five $HST$ orbits spanning 6.8~hours with 3~hours of integration time on target, which was sufficient to sample each 2.8~hour transit light curve and the baseline stellar flux before and after transit. Each observation consisted of 43 spectra with integration times of 277~s, of which 13 are in transit, four during ingress and nine near transit center.

We reduce these data using the same methods as described in previous papers from this program \citep{hui13,sing13,nik14,nik15,sing15}. Raw images were bias-, dark-, and flat-corrected with the latest version of the CALSTIS pipeline. Bad pixels flagged by CALSTIS and cosmic rays were corrected with the same routines as in \citet{nik14}. G750L spectra were fringe-corrected using a fringe flat frame obtained at the end of the observations. We use the wavelength solution determined by the $HST$ CALSTIS pipeline, which is recorded in the x1d data files.

\subsection{$Spitzer$ IRAC}

WASP-39b was observed in transit with $Spitzer$ IRAC 3.6 and 4.5~$\upmu$m channels on UT 2013 Apr 18 and UT 2013 Oct 10 respectively as part of program 90092 (P.I. D\'esert). These observations utilized the peak-up pointing mode, which places the star in the center of the targeted pixel, and included an initial 30 minute observation prior to the start of the science observation in order to allow for settling of the telescope at its new position. Science observations begin two hours before ingress and end 30 minutes after egress, capturing the entire transit and stellar baseline flux before and after transit. Each transit observation contains 8,960 subarray exposures with effective integration times of 1.92~s and total duration of 302~minutes. Data were reduced using the same methods described in \citet{knu12}, \citet{lew13}, and \citet{kam15}, including extraction of BJD$_\textrm{\scriptsize UTC}$ mid-exposure times for each image, sky background subtraction, flux-weighted centroiding to determine the position of the star in each image, and flux extraction using either a fixed or time-varying circular aperture. Flux was converted from MJy Sr$^{-1}$ to electron counts using the integration time and information in the FITS header. 

In each band, we test a fixed aperture of width 2.0 to 5.0 pixels in increments of 0.1 pixels, and time variable aperture with radius given by: 

\begin{equation}
radius = a_0 \times \sqrt{ \tilde{ \beta } } + a_1
\end{equation}

\noindent where $ \tilde{ \beta }$ is the noise pixel parameter defined in Section 2.2.2 of the IRAC handbook and is proportional to the width of the PSF, and $a_0$ and $a_1$ are scale and shift factors \citep[e.g.][]{mig05, cha08, knu12, lew13, tod13, oro14, nik15}. We test a range of $a_0$ values between 0.6 to 1.2 in increments of 0.05 while setting $a_1 = 0$, and a range of $a_1$ values from -0.8 to 0.4 pixels in increments of 0.1 while setting $a_0 = 1$. For each aperture, we remove outliers with a running median filter of 50 points and a threshold distance of 3 times the standard deviation of the fifty points, and repeat the filtering until no further points are removed. This resulted in the removal of 0.40\% and 0.39\% of the unbinned data for the chosen apertures at 3.6 and 4.5~$\upmu$m, respectively. We also trim the first hour of data for the 3.6 $\upmu$m observations, as it shows a ramp in sensitivity. The duration and strength of this ramp varies from observation to observation, but is typically strongest in the 3.6~$\upmu$m bandpass \citep[e.g.][]{knu12, lew13, zel14, kam15}. We see no evidence for a corresponding ramp in our 4.5~$\upmu$m photometry, and therefore do not trim any data in this channel.

\begin{figure}
\begin{center}
\includegraphics[scale=0.3]{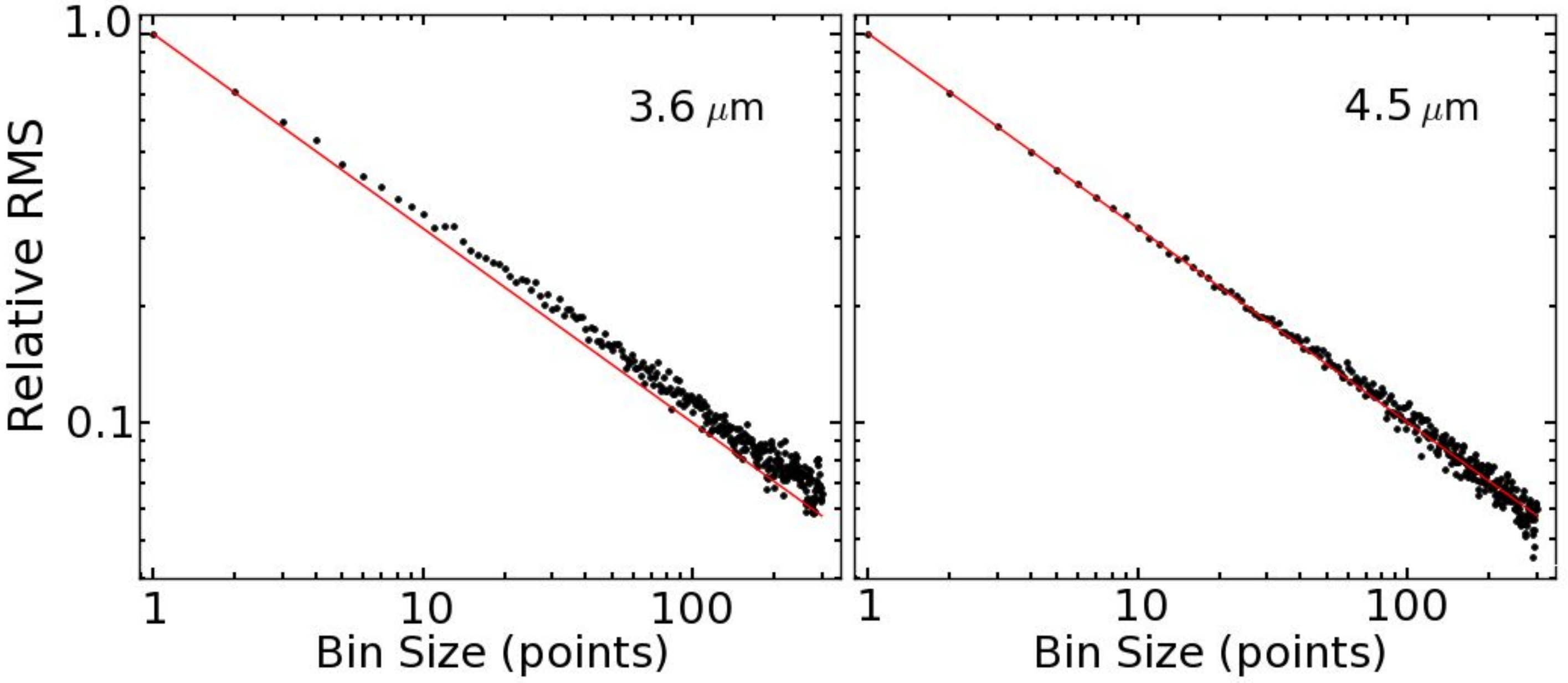}
\caption{RMS vs. bin size for the two $Spitzer$ IRAC channels. The minimal deviation from $N^{-0.5}$ shows that our final photometry for both bandpasses has a negligible amount of red noise.\label{rootN}}
\end{center}
\end{figure}

Since time-correlated or red noise dominates uncertainties for the transit parameters, solutions that minimize red noise are on the whole preferable (see Deming et al. 2015 and Kammer et al. 2015 for a more in-depth discussion of the following approach). For each aperture described in the previous paragraph we determine the best-fit instrumental and transit model by fitting to the binned time series in three minute bins; we then apply these model parameters to the unbinned light curve to calculate the unbinned residuals. We then bin these residuals and calculate the corresponding RMS as a function of bin size. For perfectly white noise we would expect this RMS to scale as $N^{-0.5}$, where $N$ is the number of points in each bin (Fig.~\ref{rootN}). We calculate the least squares difference between this ideal scaling law and the observed RMS as a function of bin size for each aperture considered.  We select our final science aperture as the one which has an unbinned RMS within 20\% of the lowest RMS aperture, and which also minimizes the red noise as quantified by our least squares metric. We find the optimal apertures are the time-variable aperture with a scale factor of 0.7 in the 3.6~$\upmu$m channel, and the fixed aperture with a radius of 2.2 pixels in the 4.5~$\upmu$m channel.

\subsection{Ground-Based Photometry}

We acquired a total of 377 nightly photometric observations of WASP-39 during 
the four observing seasons 2011-12 through 2014-15 to monitor for stellar 
activity. The observations were obtained with the Tennessee State University 
Celestron 14-inch (C14) automated imaging telescope (AIT) at Fairborn 
Observatory \citep[see, e.g.,][]{h1999,ehf2003}. Observations were made in 
the Cousins R passband with an SBIG STL-1001E CCD camera. Differential
magnitudes were computed against the mean brightness of five constant 
comparison stars in the same field. More details of our data acquisition, 
reduction procedures, and analysis techniques can be found in 
\citet{sing15}, which describes a similar analysis of the planetary-host 
star WASP-31.

Our photometric observations are summarized in Table~\ref{photometry}. The standard 
deviations of seasonal observations of WASP-39 with respect to their 
corresponding seasonal means are given in column~4, and have an average value of 0.0057~mag. Similarly, the four seasonal mean brightness values of WASP-39 given in 
column~5 scatter about their average value with a standard deviation of 
0.0018~mag with no apparent trend from year to year. Therefore, we conclude that WASP-39 is constant on both nightly and yearly timescales to the limit of our precision.

\begin{deluxetable}{ccccc}
\tablewidth{0pt}
\tablecaption{SUMMARY OF PHOTOMETRIC OBSERVATIONS FOR WASP-39\label{photometry}}
\tablehead{
\colhead{Observing} & \colhead{} & \colhead{Date Range} & 
\colhead{Sigma} & \colhead{Seasonal Mean}  \\
\colhead{Season} & \colhead{$N_{obs}$} & \colhead{(HJD$-$2,400,000)} & 
\colhead{(mag)} & \colhead{(mag)}  \\
\colhead{(1)} & \colhead{(2)} & \colhead{(3)} & 
\colhead{(4)} & \colhead{(5)}  
}
\startdata
 2011-12  &  96 & 55904--56100 & 0.0059 & $-0.5128\pm0.0006$  \\
 2012-13  &  84 & 56256--56470 & 0.0050 & $-0.5150\pm0.0005$  \\
 2013-14  & 118 & 56623--56836 & 0.0058 & $-0.5125\pm0.0005$  \\
 2014-15  &  93 & 56989--57187 & 0.0061 & $-0.5105\pm0.0006$  \\
\enddata
\end{deluxetable}

\begin{figure}
\begin{center}
\includegraphics[scale=0.38]{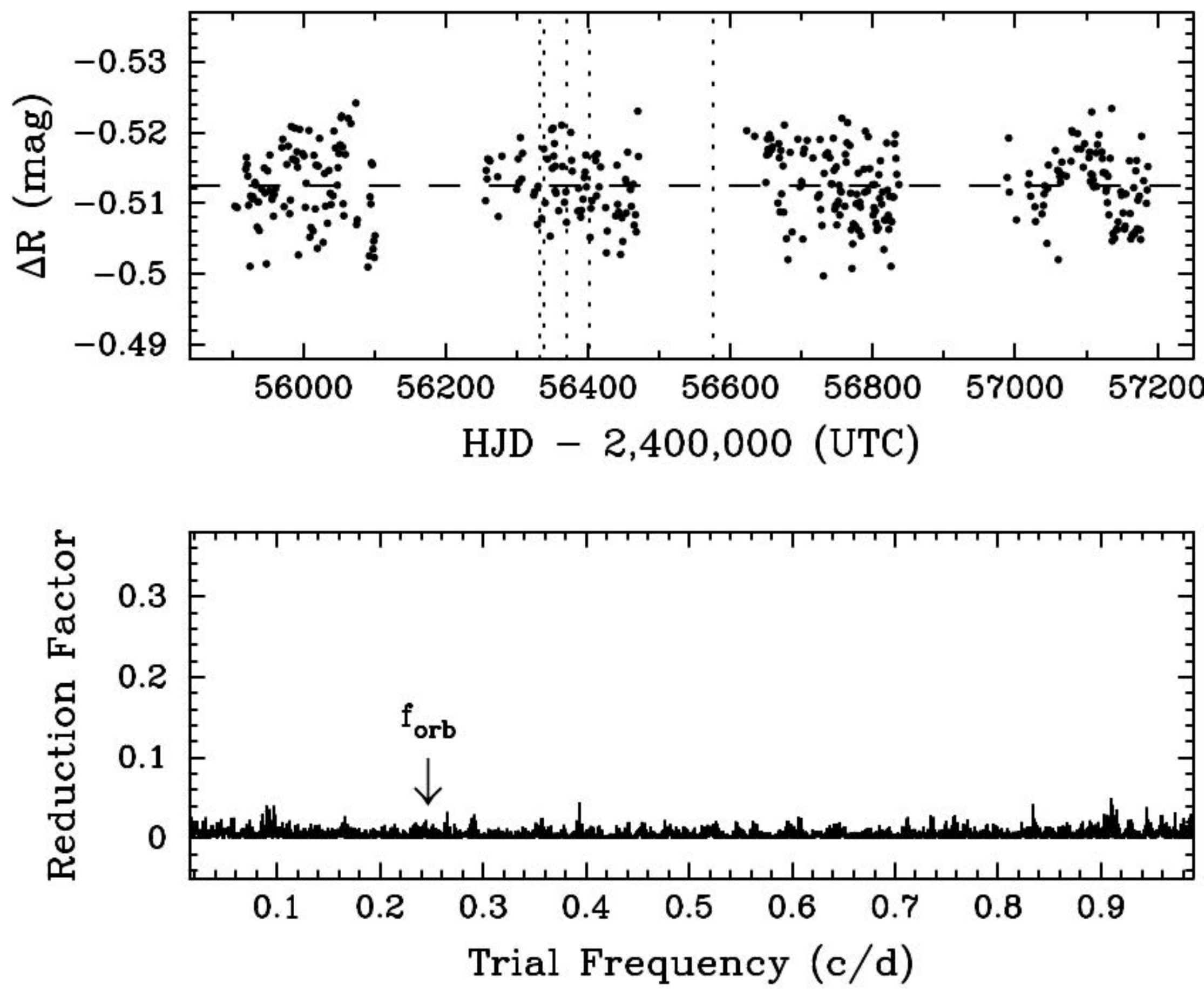}
\caption{$Top$: Four years of normalized Cousins R-band photometry of WASP-39 from the C14 automated imaging telescope at Fairborn Observatory. The observations scatter about their mean magnitude with a standard deviation of 0.0050 mag, approximately equal to C14's measurement error for a single observation. The vertical dotted lines correspond to the times of $HST$ and $Spitzer$ transit observations. $Bottom$: Frequency spectrum in cycles/day of the R-band photometry covering a period-search range of 1 to 100 days. No rotational modulation of starspots can be seen above the noise, consistent with WASP-39 being an old, late G-type dwarf. The orbital frequency is marked by an arrow and shows the absence of brightness variability on the orbital period. A least-squares sine fit gives a semi-amplitude of only $0.00050\pm0.00033$ mag.\label{activity}}
\end{center}
\end{figure}

The individual photometric observations are plotted in the top panel of 
Figure~\ref{activity}, where we have removed the 0.0018~mag scatter in the seasonal means 
by normalizing the four seasons to have the same mean.  A frequency spectrum 
over the range of 0.01 to 0.99 cycles/day, corresponding to a period range of 
1 to 100 days, is shown in the bottom panel of the figure.  No periodic 
brightness variations resulting from rotational modulation in the visibility 
of active regions and starspots can be seen above the noise level in the 
frequency spectrum.  In particular, there is no brightness variability at 
the planetary orbital frequency, marked by an arrow in the frequency spectrum. 
This corresponds to the orbital period of 4.055259~days determined by 
\citet{fae11} in their discovery paper.  A least-squares sine fit on the 
orbital period gives a semi-amplitude of only $0.00050\pm0.00033$~mag. The 
lack of rotational modulation is consistent with WASP-39 being an old, late, 
G-type dwarf \citep{fae11}. The lack of any brightness variability on the radial velocity 
period gives additional confirmation that the radial velocity variations are 
due solely to the planetary companion.

\section{Analysis}

We calculate transit light curves using the BATMAN package of \citet{kre15}. We calculate the wavelength-dependent four-parameter nonlinear limb-darkening coefficients for each bandpass using an ATLAS stellar model with an effective temperature of 5500~K and log(g) of 4.5, which is the closest match to the WASP-39 stellar parameters reported by \citet{fae11}. As in recent work \citep[e.g.][]{nik15, sing15} we choose to rely on limb-darkening coefficients determined from stellar models to reduce the number of free parameters in the fit and to avoid the degeneracies between limb-darkening parameters and transit depth. We note that the values of the limb-darkening coefficients for our $HST$ observations are not well constrained by our observations, which only sample times near the center of transit. We find consistent results using quadratic coefficients, further confirming our lack of sensitivity to the assumed limb-darkening model. The orbital period is obtained from \citet{fae11} and is also held fixed. The shape of the transit light curve then depends on four physical parameters that are fitted simultaneously for each observation: $T_c$ (transit center time), $i$ (inclination), $a/R_*$, and $R_p/R_*$, where $a$ is the semi-major orbital distance, $R_p$ is the radius of the planet, and $R_*$ is the radius of the star. Photometric time series are modeled by instrument systematic noise and transit light curves as described in the following sub-sections.

\subsection{White-Light Curves}

For the STIS observations, we first construct a white-light curve by summing the spectra across all wavelengths. We extract spectra from the images using a fixed aperture in the cross-dispersion direction. We test a range of cross-dispersion aperture widths between 1 and 30 pixels for each STIS observation and choose the one that minimizes the RMS of the residuals in the resulting white-light curves; this yielded optimal widths of 8, 10, and 9 pixels for the three observations respectively. We use these white-light curves to determine optimal values for the wavelength-independent parameters (transit ceter time $T_c$, inclination $i$, normalized semi-major axis $a/R_*$). As in previous studies \citep[e.g.][]{hui13, v-m13, sing13, nik14, nik15, sing15}, we find that the STIS sensitivity variations are adequately described by a linear function in time across all $HST$ orbits (two parameters), and a fourth-order polynomial in time phased to each $HST$ orbit (four parameters), totaling six instrumental noise parameters per STIS observation. We test a range from 2nd order to 6th order polynomials for the orbital-phased systematic noise, and found that fourth order optimized the BIC. Consistent with previous work, we see possible evidence for differing systematic noise in the first $HST$ orbit of each observation, which is thought to be caused by thermal relaxation after a new pointing \citep{v-m13}. We therefore exclude the first orbit of each observation in our analysis. For the $Spitzer$ IRAC light curves we utilize the pixel-level decorrelation technique \citep{dem15} with a grid of nine pixels centered on the position of the star. This results in a total of 10 free parameters for each $Spitzer$ light curve, including nine pixel weighting coefficients and a constant term.

As a result of gaps in time coverage due to $HST$'s low-earth orbit, we find that the STIS observations alone do not provide strong constraints on the planet's orbital inclination and $a/R_*$. Fortunately, our $Spitzer$ IRAC observations span the entire transit with no gaps and are obtained at longer wavelengths where the effects of stellar limb-darkening are minimal. We therefore perform a simultaneous fit with STIS white-light curves and IRAC 3.6 and 4.5~$\upmu$m light curves, in which inclination and $a/R_*$ are common for all observations, but $T_c$, $R_p/R_*$, and instrumental noise parameters (see previous paragraph for a description of this model for $HST$ and equivalent $Spitzer$ intrumental noise model) are unique for each transit observation, resulting in a total of 50 free parameters to model 290 photometric points in the global fit for the combined $HST$ and $Spitzer$ light curves. The instrumental parameters are fit simultaneously with the transit parameters. We report a common $R_p/R_*$ value for the two STIS G430L observations, which is valid assuming negligible contributions from stellar variability, consistent with the results of the stellar activity monitoring described previously. 

We determine the best-fit model parameters in the global fit using an MCMC analysis, implemented using the EMCEE python routine of \citet{dfm13}. Achieving accurate posterior distributions with MCMC requires accurate uncertainties for the data being modeled. To ensure this, we first fit the white-light curves with a simple least squares minimization and calculate the standard deviation of the residuals for each individual $HST$ and $Spitzer$ transit. We then set the photometric uncertainties within each transit to the corresponding standard deviation from this fit. We assume uniform priors on all fit parameters, and limit the inclination to values less than 90 degrees. We run an initial chain with 50,000 steps in order to determine the optimal step sizes, and then re-run a longer chain with $10^6$ total steps and 500 walkers in order to calculate our final posterior probability distributions. We check convergence by dividing our chain into 4 sub-segments and re-calculating the median value of the chain for each fit parameter. We find that these median values are all within 0.2\% of the nominal values from the global chain. The posterior probability distributions from our fits are effectively Gaussian for all of the astrophysical parameters aside from inclination and $a/R_*$, which are known to be correlated in transit fits. Our instrumental noise parameters for a given light curve also show some correlations with each other, but appear to be uncorrelated with our transit shape parameters. The reported results are the means of the MCMC posterior distributions and are shown in Table~\ref{results1}. Parameter uncertainties are determined by marginalizing over the MCMC chain for each parameter to find its posterior probability distribution and calculate the corresponding 68\% confidence interval. 

\begin{figure*}
\includegraphics[scale=0.49]{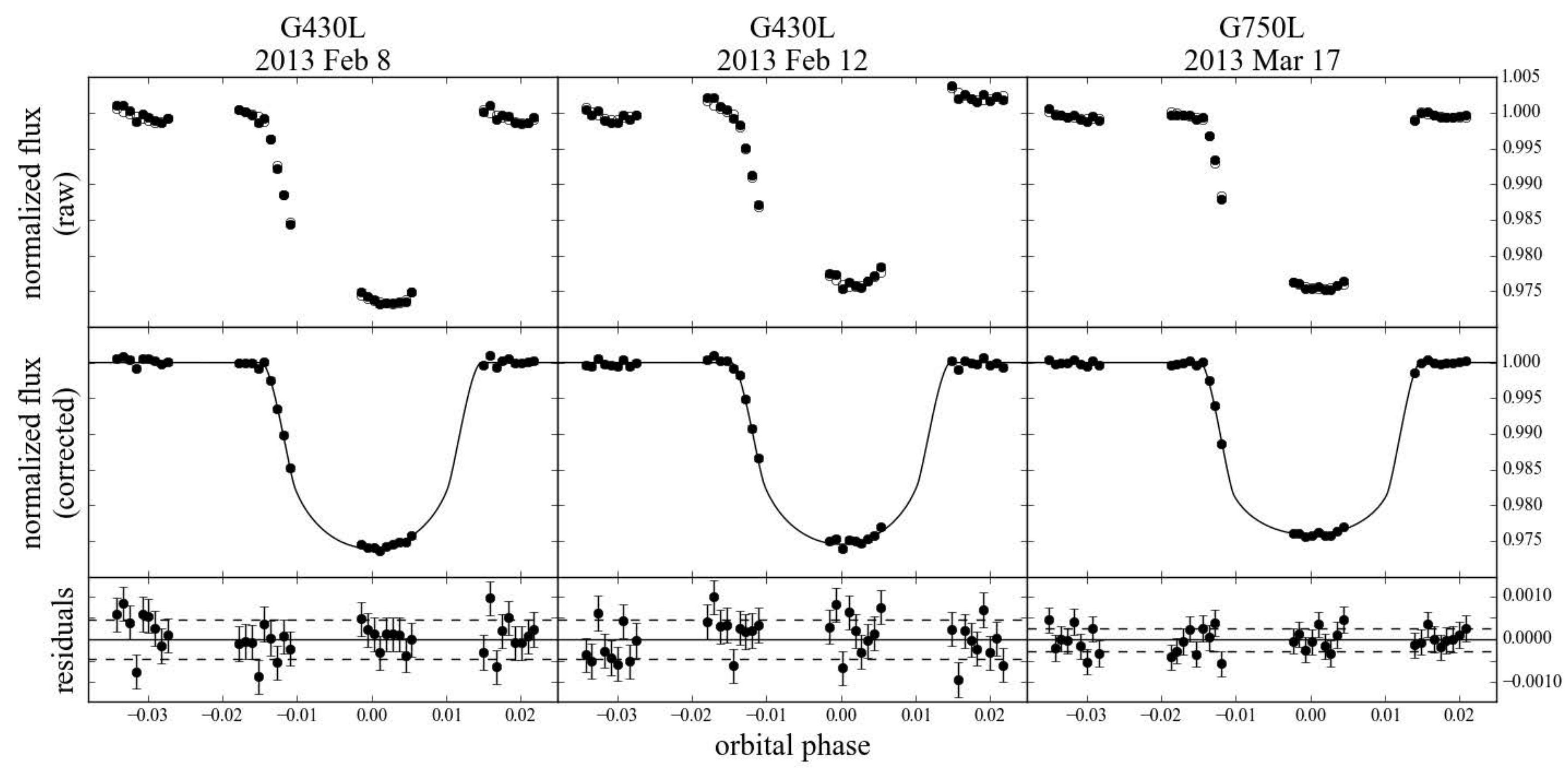}
\caption{$HST$ STIS raw (top) and corrected (middle) white-light curves, where we have divided out the best-fit instrumental noise model for each visit. Open circles in the top panel show the model points. The bottom panel shows the residuals after removing both the instrumental model and transit light curves. Error bars are the standard deviation of these residuals. Dashed lines show one standard deviation.\label{stis}}
\end{figure*}

\begin{figure*}
\begin{center}
\includegraphics[scale=0.5]{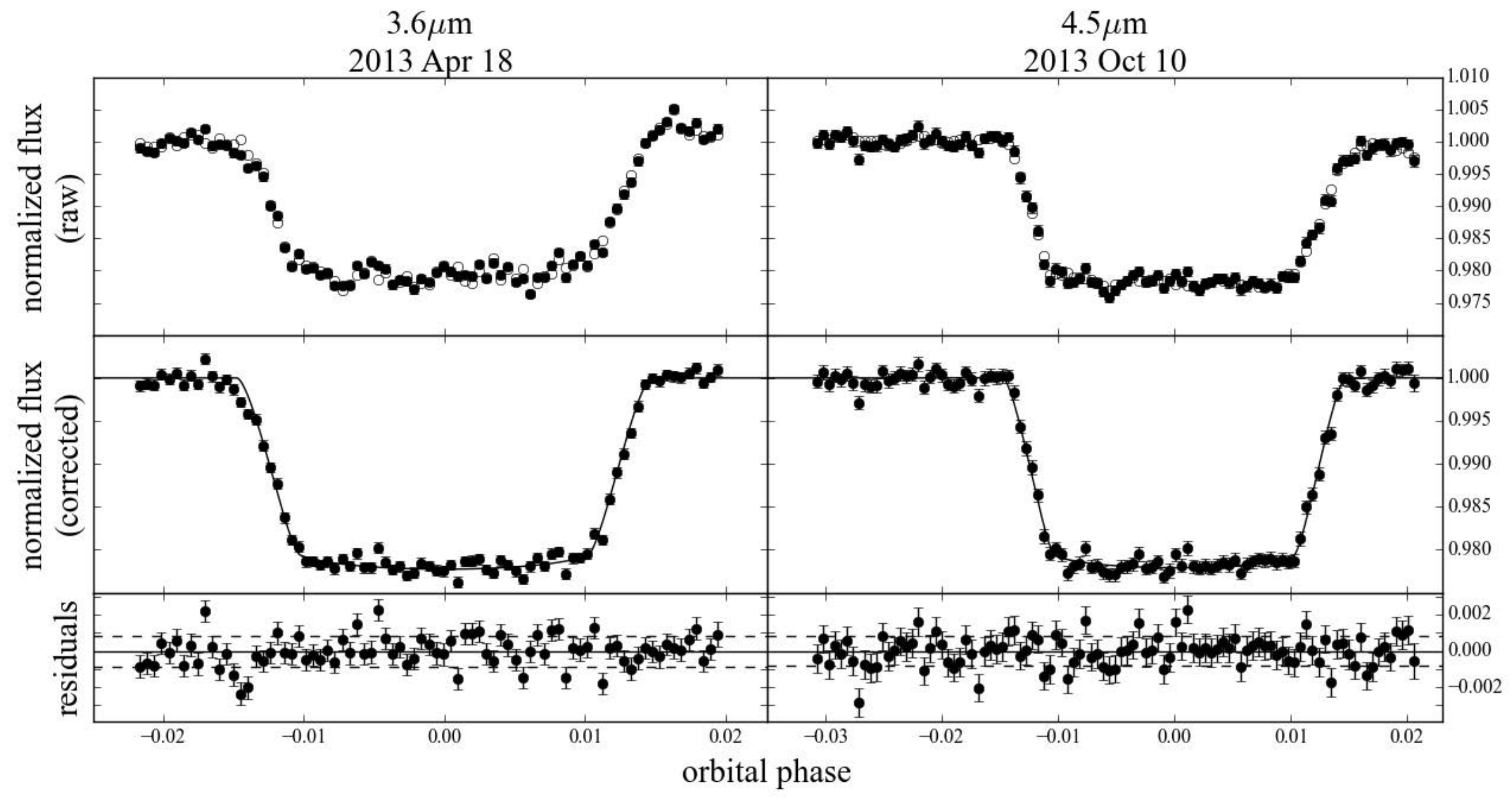}
\caption{$Spitzer$ IRAC 3.6- and 4.5-$\upmu$m lightcurves for raw (top) and corrected (middle) white-light curves, where we have binned the data in time to three-minute intervals and divided out the best-fit instrumental noise model for each visit. Open circles in the top panel show the model points. The bottom panel shows the residuals after removing both the instrumental model and transit light curves. Error bars are the standard deviation of these residuals. Dashed lines show one standard deviation.\label{spitzer}}
\end{center}
\end{figure*}

\subsection{Spectral Light Curves}

\begin{figure*}
\begin{center}
\includegraphics[scale=0.51]{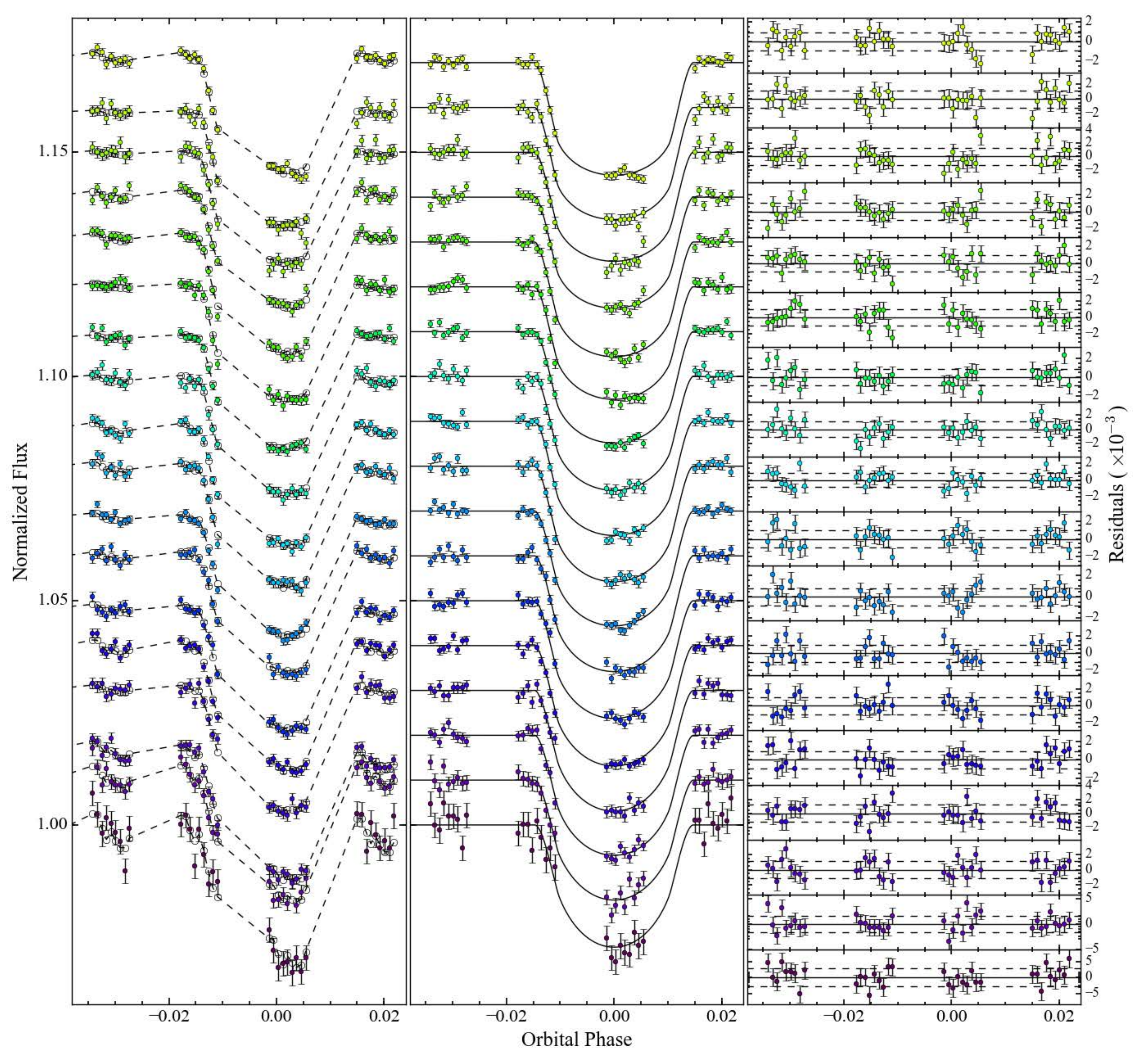}
\caption{$HST$ STIS G430L spectral light curves for the observations on UT 2013 Feb 8, raw (left) and corrected (middle) where we have divided out the best-fit instrumental noise model for each visit. Color corresponds to wavelength, with the shortest wavelengths plotted in purple and the longest wavelengths plotted in light green. Light curves are offset vertically by multiples of 0.01. Open circles in the left panel show the model points. The right panel shows the residuals after removing both the instrumental model and transit light curves. Dashed lines show one standard deviation. \label{binned13}}
\end{center}
\end{figure*}

\begin{figure*}
\begin{center}
\includegraphics[scale=0.51]{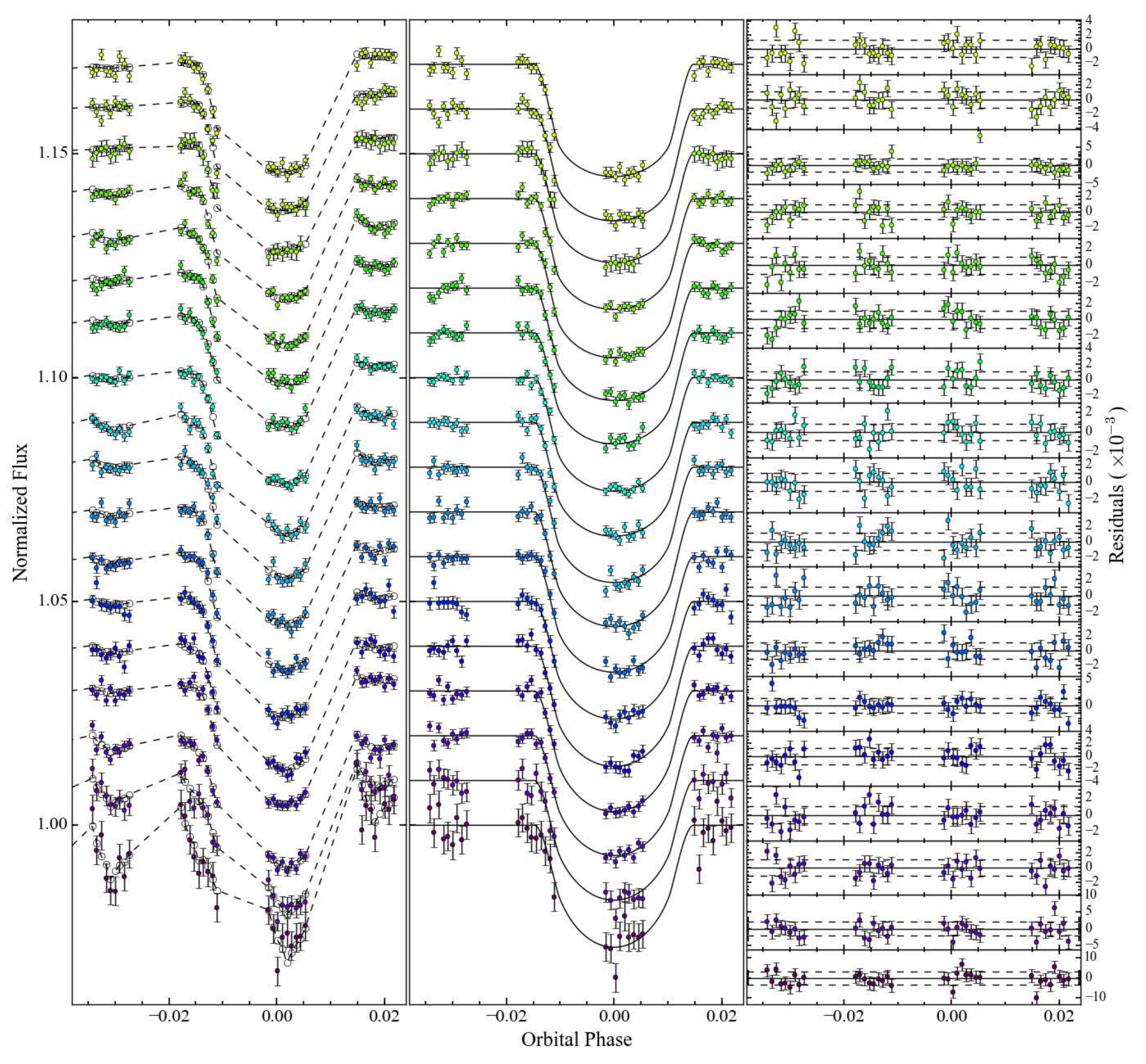}
\caption{Same description as in Fig.~\ref{binned13}, for observations on UT 2013 Feb 12.\label{binned14}}
\end{center}
\end{figure*}

\begin{figure*}
\begin{center}
\includegraphics[scale=0.51]{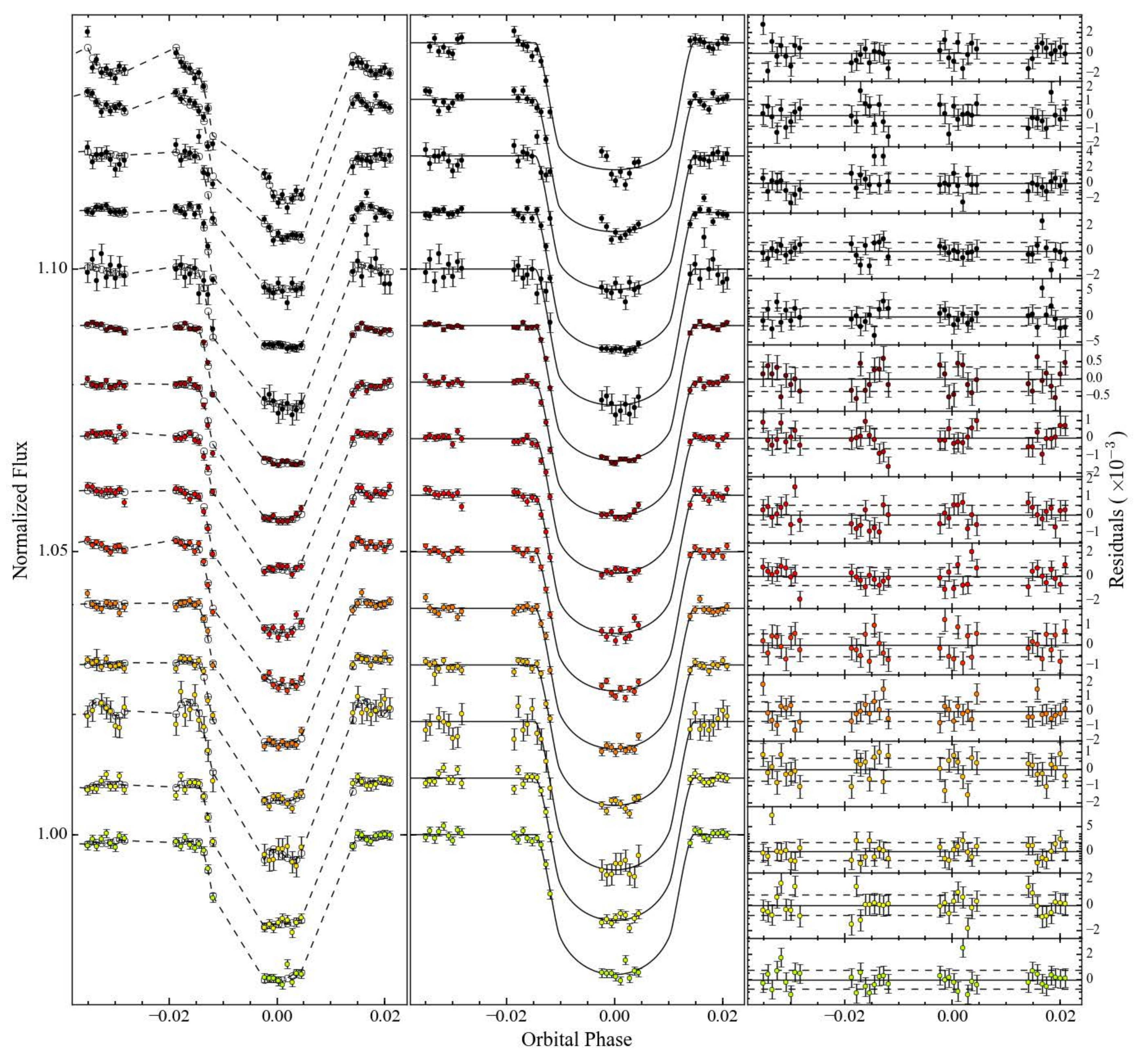}
\caption{Same description as in Fig.~\ref{binned13}, \ref{binned14} for observations with the G750L grism on UT 2013 Mar 17.\label{binned23}}
\end{center}
\end{figure*}

The transmission spectrum is obtained by grouping the STIS spectra into 35 bins in the dispersion direction with wavelength ranges shown in Table~\ref{results2}, which are much coarser than the instrument resolution of 0.27 and 0.49 nm/pixel in the G430L and G750L grisms, respectively. Each binned time series yields a unique spectral light curve (Fig.~\ref{binned13}, \ref{binned14}, \ref{binned23}). We select the wavelength range of the spectral bins to optimally balance resolution and S/N, and to distinguish sodium and potassium features. We calculate four parameter nonlinear limb-darkening parameters unique to each bandpass. We fit the spectral light curves with the transit model with $R_p/R_*$ as a free parameter, but fix the wavelength-independent parameters $T_c$, inclination, and $a/R_*$ to the values obtained from the global white-light curve fit. We assume a common value for $R_p/R_*$ in each bandpass for the two observations with the G430L grism and fit both observations simultaneously, as in the white-light curve fit. We note that fitting for the two G430L observations separately yielded results that were consistent at approximately the 1.4$\sigma$ level in both the spectral and white-light fits, with a small constant offset possibly related to the instrumental noise model for each visit. We find that the transmission spectrum from the joint fit to both G430L visits is in good agreement with the G750L grism data in the region where the wavelengths overlap, and report the values from this joint fit in Table~\ref{results2}.

We model the instrumental systematic noise in the spectral light curves using both a parametric model and a common-mode model as discussed below. In both cases we fit the transit model simultaneously with the instrumental noise model. The parametric model is the same as described previously for the white-light curves, but with a unique fit for the light curve in each individual wavelength bin. This model includes a fourth order polynomial phased to the $HST$ orbit, which accounts for the breathing of the telescope, and a linear function of time to account for longer-term trends across spacecraft orbits. In the common-mode model we replace this fourth order polynomial with an empirical noise model based on the residuals from the white-light fit after the transit and best-fit linear trend have been removed. We then fit for a new linear function of time in each bandpass, and allow the amplitude of the white-light residual noise model to vary as a free parameter in our fits.

Both instrumental noise models give equivalent results for the transmission spectrum, with the most significant difference in the shortest wavelength bin of the G430L grism where the grating efficiency is lower and the spectral light curve correspondingly noisier. Here the common-mode method favors a higher $R_p/R_*$ value than the parametric model by approximately 1$\sigma$, and is in better agreement with the rest of the transmission spectrum. This agrees well with results from previous studies \citep[e.g.][]{sing13, sing15}, which found that the common-mode method performs better than polynomial noise models in bands with higher intrinsic noise levels. We therefore report the values from this common-mode model in Table~\ref{results2}. In the G750L grism we find that the shape of the instrumental noise varies as a function of wavelength, and we therefore obtain a better fit to these data using the polynomial model for the instrumental noise. We note a similar trend in the G750L observations of HAT-P-1b, where we also preferred a polynomial noise model \citep{nik14}. As with the white-light curves, we use a MCMC fit to determine our best-fit parameters and corresponding uncertainties for each bandpass. The common-mode model for the wavelength-dependent G430L light curves has seven free parameters: a common $R_p/R_*$ for the two G430L observations, individual amplitude factors for each G430L observation, and individual linear sensitivity intercept and slope values for each observation. The parametric model for the wavelength-dependent G750L light curves has seven free parameters: $R_p/R_*$, linear sensitivity intercept and slope, and four polynomial coefficients. We find that the posteriors for our wavelength-dependent $R_p/R_*$ values are all Gaussian, and list these values along with their corresponding uncertainties in Table~\ref{results2}.

\begin{table*}
\begin{center}
\caption{Transit Parameters from Global Fit.\label{results1}}
\begin{tabular}{lllll}
\tableline
\tableline
\\
Instrument & Date & Wavelength ($\textrm{\AA}$) & Parameter & Value\\
\tableline
\\
STIS G430L & 2013 Feb 8, 12 & 2896 - 5706 & Planet radius ($R_p / R_*$) & 0.14463 $\pm$ 0.00069 \\
STIS G430L & 2013 Feb 8 & 2896 - 5706 & Transit center BJD$_\textrm{\scriptsize TDB}$ (days) & $2456332.45857^{+0.00035}_{-0.00031}$ \\
STIS G430L & 2013 Feb 12 & 2896 - 5706 & Transit center BJD$_\textrm{\scriptsize TDB}$ (days) & $2456336.51323 ^{+0.00033}_{-0.00034}$ \\
\\
STIS G750L & 2013 Mar 17 & 5259 - 10251 & Planet radius ($R_p / R_*$) & 0.14452 $\pm$ 0.00059 \\
STIS G750L & 2013 Mar 17 & 5259 - 10251 & Transit center BJD$_\textrm{\scriptsize TDB}$ (days) & $2456368.95549 ^{+0.00024}_{-0.00026}$ \\
\\
IRAC 3.6-$\upmu$m & 2013 Apr 18 & 31618 - 39284 & Planet radius ($R_p / R_*$) & 0.14513 $\pm$ 0.00148 \\
IRAC 3.6-$\upmu$m & 2013 Apr 18 & 31618 - 39284 & Transit center BJD$_\textrm{\scriptsize TDB}$ (days) & $2456401.39733 ^{+0.00022}_{-0.00036}$ \\
\\
IRAC 4.5-$\upmu$m & 2013 Oct 10 & 39735 - 50198 & Planet radius ($R_p / R_*$) & 0.14591 $\pm$ 0.00096 \\
IRAC 4.5-$\upmu$m & 2013 Oct 10 & 39735 - 50198 & Transit center BJD$_\textrm{\scriptsize TDB}$ (days) & $2456575.77465 ^{+0.00024}_{-0.00021}$ \\
\\
Combined & & & Semi-major axis ($a / R_*$) & 11.55 $\pm$ 0.13 \\
Combined & & & Orbital inclination ($^{\circ}$) & 87.93 $\pm$ 0.14 \\
\tableline\tableline
\end{tabular}
\end{center}
\end{table*}

\begin{table}
\begin{center}
\caption{STIS Transmission Spectrum Results.\label{results2}}
\begin{tabular}{cc}
\tableline\tableline
\\
Wavelength ($\textrm{\AA}$) & $R_p/R_*$\\
\tableline
\\
\sidehead{STIS G430L}
2900 - 3700 & 0.14429 $\pm$ 0.00230\\
3700 - 3950 & 0.14408 $\pm$ 0.00143\\
3950 - 4113 & 0.14467 $\pm$ 0.00089\\
4113 - 4250 & 0.14500 $\pm$ 0.00088\\
4250 - 4400 & 0.14653 $\pm$ 0.00087\\
4400 - 4500 & 0.14491 $\pm$ 0.00093\\
4500 - 4600 & 0.14396 $\pm$ 0.00085\\
4600 - 4700 & 0.14376 $\pm$ 0.00073\\
4700 - 4800 & 0.14447 $\pm$ 0.00086\\
4800 - 4900 & 0.14461 $\pm$ 0.00076\\
4900 - 5000 & 0.14381 $\pm$ 0.00081\\
5000 - 5100 & 0.14301 $\pm$ 0.00081\\
5100 - 5200 & 0.14361 $\pm$ 0.00089\\
5200 - 5300 & 0.14541 $\pm$ 0.00077\\
5300 - 5400 & 0.14330 $\pm$ 0.00081\\
5400 - 5500 & 0.14309 $\pm$ 0.00122\\
5500 - 5600 & 0.14437 $\pm$ 0.00098\\
5600 - 5700 & 0.14558 $\pm$ 0.00083\\
\sidehead{STIS G750L}
5500 - 5650 & 0.14381 $\pm$ 0.00115\\
5650 - 5880 & 0.14550 $\pm$ 0.00126\\
5880 - 5910 & 0.14890 $\pm$ 0.00282\\
5910 - 6060 & 0.14496 $\pm$ 0.00109\\
6060 - 6300 & 0.14531 $\pm$ 0.00111\\
6300 - 6450 & 0.14491 $\pm$ 0.00091\\
6450 - 6600 & 0.14520 $\pm$ 0.00121\\
6600 - 6800 & 0.14315 $\pm$ 0.00091\\
6800 - 7100 & 0.14301 $\pm$ 0.00092\\
7100 - 7650 & 0.14423 $\pm$ 0.00057\\
7650 - 7710 & 0.14532 $\pm$ 0.00283\\
7710 - 8100 & 0.14604 $\pm$ 0.00109\\
8100 - 8500 & 0.14451 $\pm$ 0.00200\\
8500 - 9000 & 0.14425 $\pm$ 0.00122\\
9000 - 10250 & 0.14179 $\pm$ 0.00161\\
\tableline\tableline
\\
\end{tabular}
\end{center}
\end{table}

\subsection{White-Light Transit Parameters and Updated Orbital Ephemeris}

\begin{figure}
\begin{center}
\includegraphics[scale=0.43]{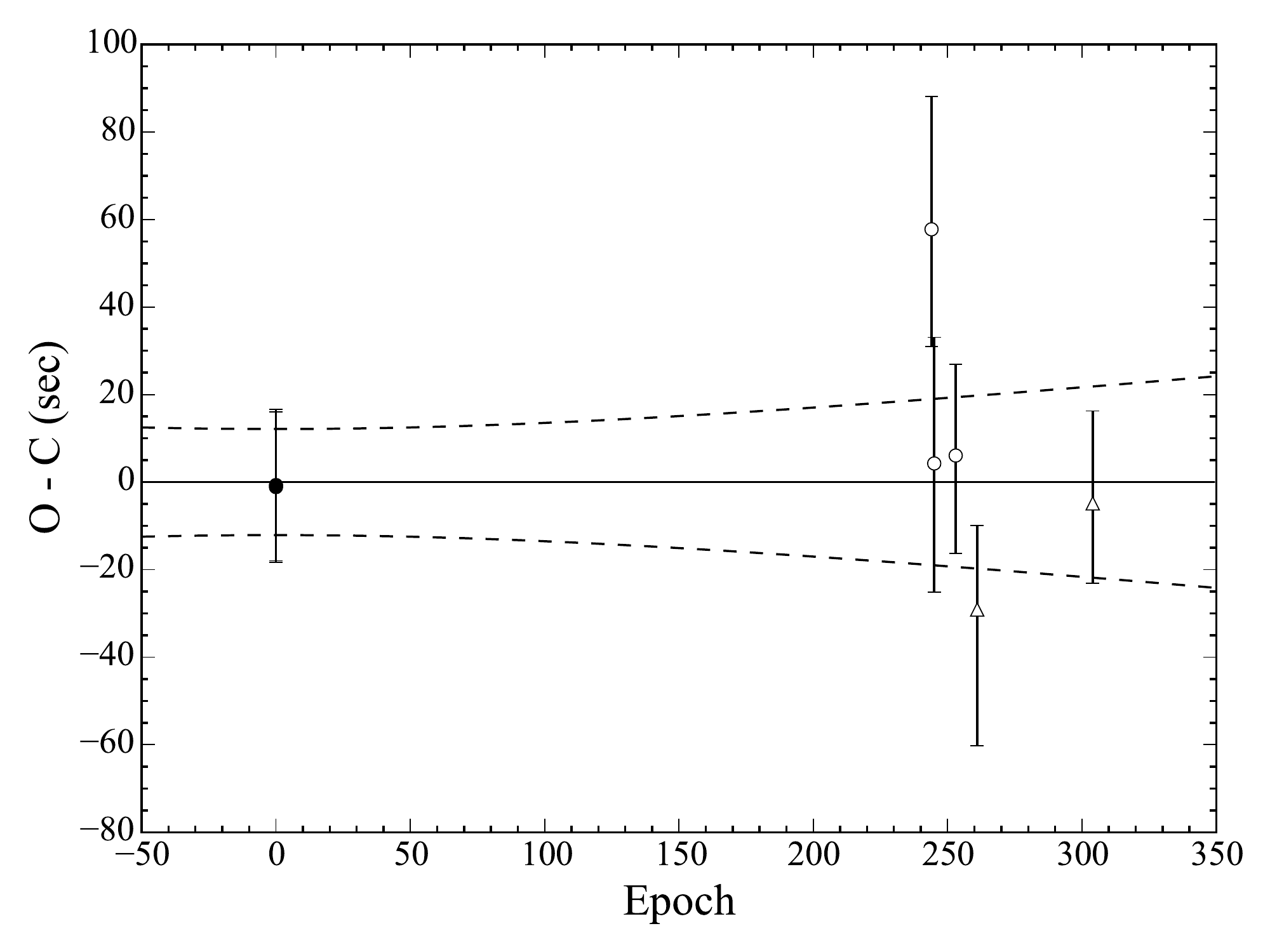}
\caption{Observed - computed times of transit center for $HST$ (open circles), $Spitzer$ (open triangles), and the zero epoch times reported by \citet{ric15, fae11} (filled circles, nearly overlapping). Dotted lines show the 1$\sigma$ uncertainty. All times have been converted to BJD$_\textrm{\scriptsize TDB}$ following \citet{eas10}.\label{oc}}
\end{center}
\end{figure}

From the global fit combining STIS and IRAC observations, we find common values for inclination and semi-major axis of $i=87.93\pm 0.14$ and $a/R_*=11.55\pm 0.13$, in agreement with the values reported by \citet{fae11} and \citet{ric15}. The transit times calculated from our combined fit (Table~\ref{results1}) and the result from \citet{fae11} and \citet{ric15} are fit as a function of period $P$ and epoch $E$,

\begin{equation}
T(E) = T(0) + E \times P
\end{equation}

\noindent where the initial transit epoch $T(0)$ is chosen to remain consistent with previous work (Fig.~\ref{oc}). We convert all transit times to BJD$_\textrm{\scriptsize TDB}$ following \citet{eas10}. We find an initial transit epoch $2455342.9696\pm0.00014$~BJD$_\textrm{\scriptsize TDB}$ and period $4.05527999\pm7.0\times10^{-7}$~days. \citet{ric15} previously reported a period of $4.0552947\pm9.65\times10^{-7}$, which disagreed with the value found by \citet{fae11} of $4.055259\pm8\times~10^{-6}$ at the 4.4$\sigma$ level. Our best-fit period is intermediate between these values. We differ from the value reported by \citet{fae11} at the 2.6$\sigma$ level, and from the value in \citet{ric15} by 12$\sigma$. Although transit timing variations might explain these discrepant results, we find no evidence for timing variations in the transits analyzed here, and the previous studies do not report individual transit times.

\subsection{Transmission Spectrum Models}

We find that the transmission spectrum is well-matched by clear-atmosphere models including broad line absorption wings from both Na and K. Although weak contributions from hazes or clouds are also consistent, additional sources of opacity are not required to describe the observed transmission spectrum. 

We use the Rayleigh slope shortward of 520~nm to empirically measure the temperature of the planet's atmosphere at the day-night terminator. This slope is determined by:

\begin{equation}
\alpha T = \frac{\mu g}{k} \frac{d(R_p/R_*)}{d~ln(\lambda)}
\end{equation}

\noindent \citep{lec08} where $\mu$ is the mean molecular mass, $g$ is the surface gravity, $k$ is the Boltzman constant, $T$ is temperature, and $\alpha$ is the index that defines the wavelength-dependence of the scattering cross-section, $\sigma/\sigma_0=(\lambda/\lambda_0)^\alpha$. We set $\alpha$ equal to -4 for Rayleigh scattering and fix the surface gravity to the value reported in \citet{fae11} of 407 cm s$^{-2}$. Using this expression we find a best-fit terminator temperature of 940$\pm$470~K, in good agreement with the predicted zero-albedo terminator equilibrium temperature $T_{eq}$ = 1120$\pm$30 \citep{fae11}.

\begin{figure*}
\begin{center}
\includegraphics[scale=0.29]{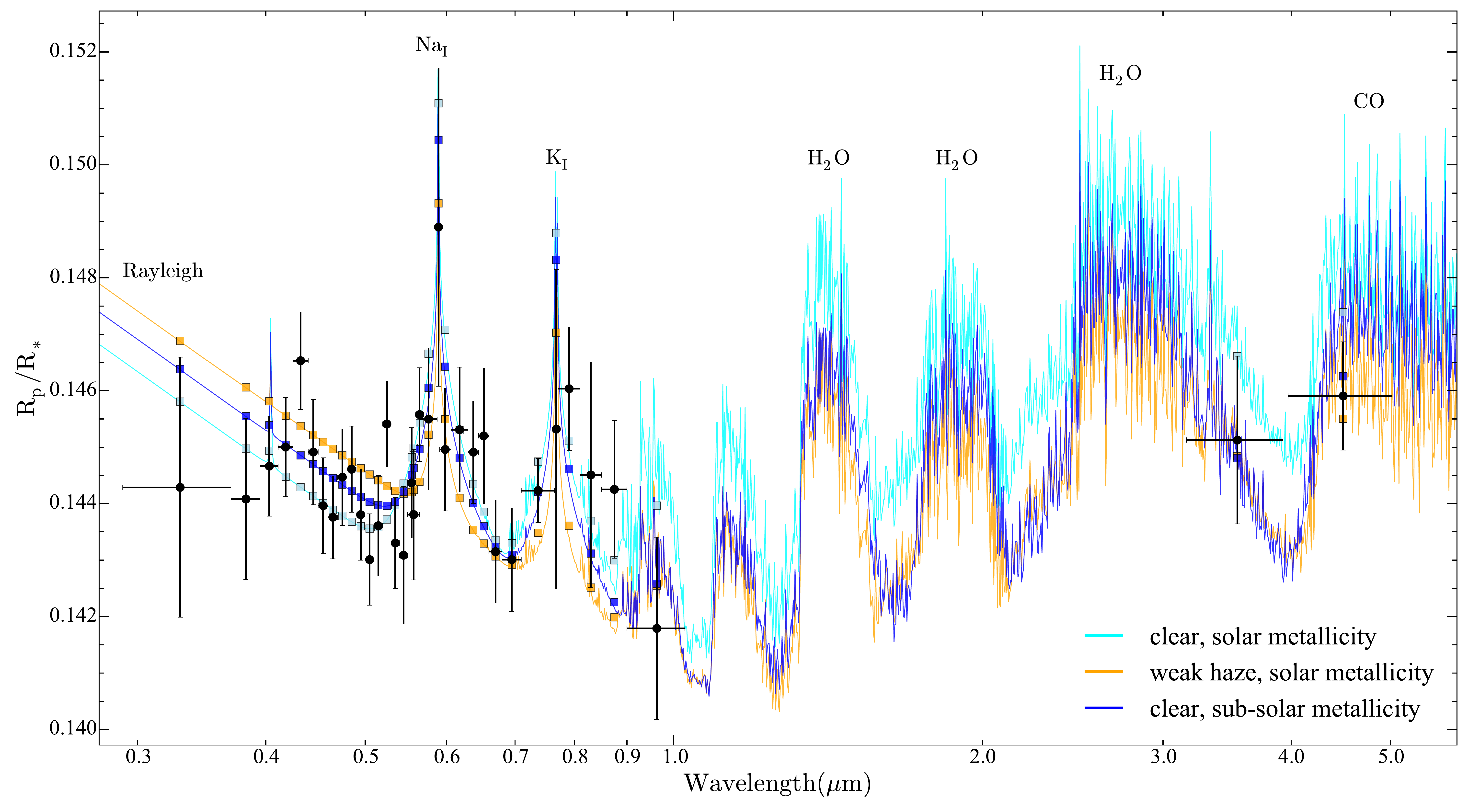}
\caption{Measured transmission spectrum and forward models scaled to WASP-39b for Clear solar metallicity (cyan), Solar metallicity with weak haze (orange), and Clear sub-solar (0.1$\times$) metallicity (blue). Squares show the model binned to the data wavelengths. Each model is consistent with the transmission spectrum to within the uncertainties.\label{fortneymodels}}
\end{center}
\end{figure*}

Figure~\ref{fortneymodels} shows the measured transmission spectrum with several atmosphere models overplotted for comparison, and Table~\ref{fits} gives the corresponding $\chi^2$ and Bayesian Information Criterion (BIC) values for each model. We compare the transmission spectrum over the entire STIS (290-1025~nm) and IRAC (3.6, 4.5~$\upmu$m) wavelength range (Fig.~\ref{fortneymodels}) with forward models with a temperature of 1000~K from \citet{for10}, which we scale to match the measured surface gravity and radius of WASP-39b. We consider models for solar metallicity, solar metallicity with weak haze characterized by enhanced (10$\times$) Rayleigh scattering, and a sub-solar (0.1$\times$) metallicity (new run of the Fortney et al. 2010 model). We fit for a vertical offset as the only free parameter, totaling 35 data points and 34 degrees of freedom for each fit.

We find that the observed transmission spectrum is best matched by the sub-solar metallicity model (blue), which has a $\chi^2$ of 28.0. It is also consistent with the clear solar metallicity model (cyan), which has a $\chi^2$ of 35.8, and the weak haze model (orange), which has a $\chi^2$ of 41.8. Strong hazes such as those seen in HD~189733b or WASP-6b \citep{pon13, nik15} are ruled out; 100$\times$ and 1000$\times$ enhanced Rayleigh models produce poor fits with $\chi^2$ values of 80.3 and 130, respectively. We note that these models show a degeneracy between the effects of depleted abundances and enhanced scattering. As a justification for the subsolar metallicity model, we note that there are two possible routes by which we might deplete the Na and K abundances in WASP-39b's upper atmosphere. If the night side or deep atmosphere temperature-pressure profiles cross the condensation curves for these elements, then circulation of gas between these regions and the upper atmosphere would naturally result in their depletion \citep{kat16}. Alternatively, the planet might also have accreted a metal-poor atmosphere from the primordial gas disk; although the host star may be moderately metal-poor ([Fe/H] = -0.12$\pm$0.1), our 0.1$\times$ solar metallicity model would still require some additional segregation of heavy elements in the planet core.

\begin{figure}
\begin{center}
\includegraphics[scale=0.32]{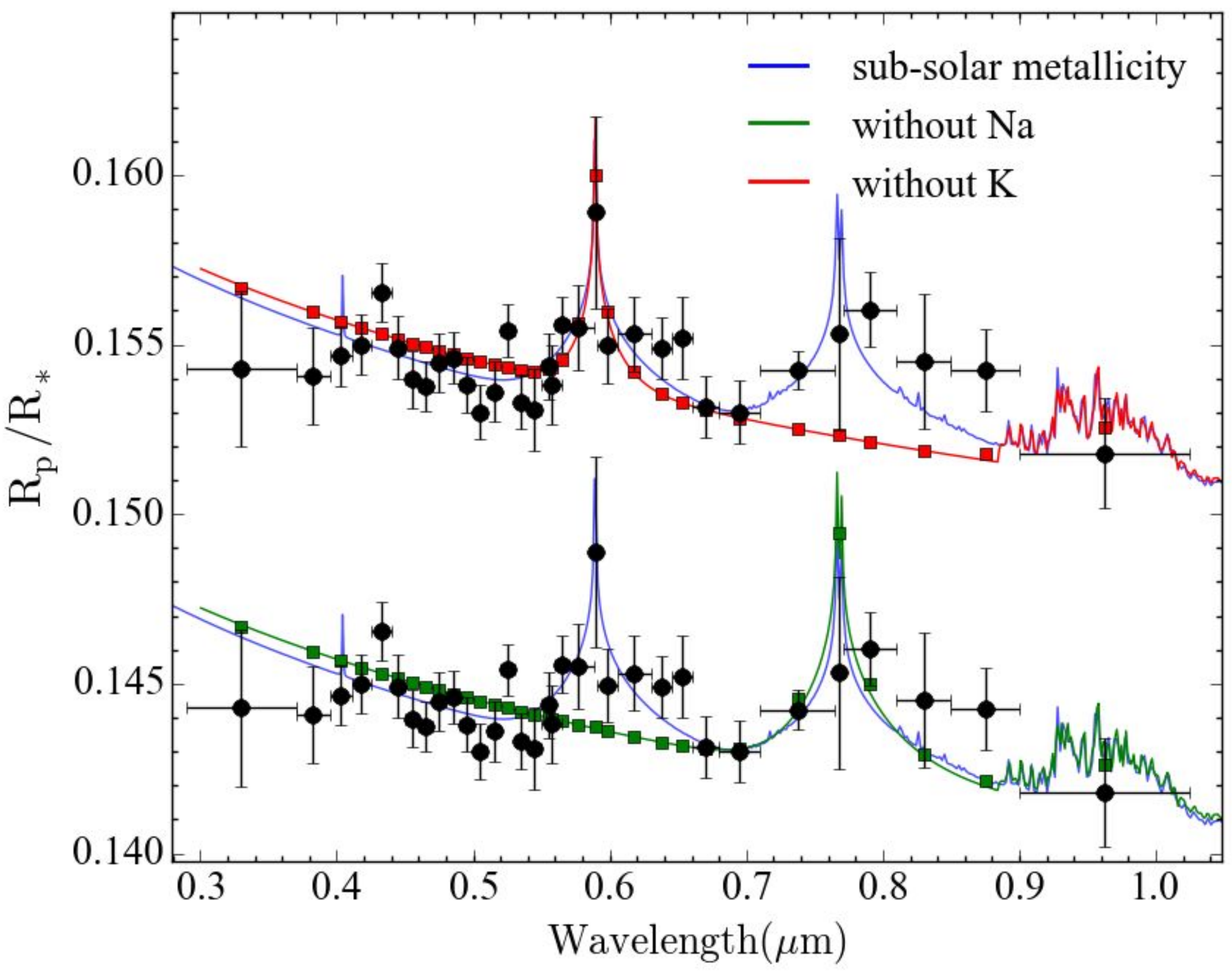}
\caption{Comparison of analytical models with no Na (green) or no K (red) to the measured transmission spectrum (black filled circles). We also plot the 0.1$\times$ solar metallicity model (blue) for comparison. The top spectra have been offset vertically by 0.01 $R_p / R_*$.\label{analyticalmodels}}
\end{center}
\end{figure}

We also fit the optical transmission spectrum with analytical models (Fig.~\ref{analyticalmodels}), similar to \citet{sing15}. The analytical model allows us to test the significance of the Na and K line detections by removing these species from the fit individually and calculating the difference in $\chi^2$. We calculate the optical transmission spectrum following \citet{lec08} with the expression:

\begin{equation}
z(\lambda) = H~ln\left( \frac{\epsilon_{abs}P_{ref}\sigma({\lambda})}{\tau_{eq}}\sqrt{\frac{2\pi R_p}{kT\mu g}}\right)
\end{equation}

\noindent where $z(\lambda)$ is the transmission altitude, $\epsilon_{abs}$ is the abundance of dominating absorbing species as number fraction relative to total number of molecules, $P_{ref}$ is the pressure at the reference altitude, $\sigma(\lambda)$ is the absorption cross-section, and $H = kT/\mu g$ is the atmosphere scale height. As we have not spectrally identified IR molecular features, such as H$_2$O, we do not explicitly fit for their abundances in the analytic model, but rather we utilize the sub-solar metallicity model from the previous section to describe the shape of the transmission spectrum longward of 885~nm, where we shift this model in order to match the value of the analytical model at 885~nm. The analytical model includes 35 data points and 31 d.o.f. with four free parameters: Na abundance, K abundance, optical baseline radius $z_0$, and infrared baseline radius $z_{ir}$ (the fitted infrared baseline radius is a proxy for the unknown molecular abundances, determined from the transmission spectrum in the infrared). We also fit the analytical model while fixing either the Na abundance or K abundance to zero, giving three free parameters and 32 d.o.f. in these fits. Fitting for both Na abundance and K abundance as free parameters simultaneously gives $\chi^2=33.4$, with ln$(\epsilon_\textrm{\scriptsize Na})=-15.6\pm0.6$ (0.1$\times$ solar), ln($\epsilon_\textrm{\scriptsize K})=-17.6\pm0.5$ (0.2$\times$ solar), in reference to the solar abundance values from \citet{asp09}. Fixing either Na abundance or K abundance to zero (Figure~\ref{analyticalmodels} green and red lines) yields $\chi^2$ values of 47.4 or 57.7 respectively. A likelihood ratio test of nested models gives a detection significance of 3.7$\sigma$ for Na and 4.9$\sigma$ for K. Although we conclude that both species are clearly present in this planet's transmission spectrum, we note that their abundances can increase by an order of magnitude depending on whether or not we allow for the possibility of a weak haze at the shortest wavelengths. Thus, the Na and K abundances in the clear-atmosphere scenario, where the blue scattering slope is due to molecular hydrogen, can be considered lower limits. However the Na/K abundance ratio is not dependent on P$_\textrm{\scriptsize ref}$ \citep{sing15}, and can be accurately measured independent of whether the atmosphere is clear or contains a weak haze. We find a Na to K abundance ratio for WASP-39b that is 45\%$\pm$31\% of the solar value, in reference to the solar abundance values from \citet{asp09}. 

\begin{table}
\begin{center}
\caption{Model fit statistics.\label{fits}}
\begin{tabular}{lccc}
\tableline\tableline
\\
Model & $N$, d.o.f. & $\chi^2$ & BIC\tablenotemark{a}\\
\tableline
\\
0.1$\times$ solar metallicity, clear & 35, 34 & 28.0 & 31.6\\
solar metallicity, clear & 35, 34 & 35.8 & 39.4\\
solar metallicity, 10$\times$ scattering & 35, 34 & 41.8 & 45.4\\
solar metallicity, 100$\times$ scattering & 35, 34 & 80.3 & 83.9\\
solar metallicity, 1000$\times$ scattering & 35, 34 & 130 & 134\\ 
\\
\tableline
\\
analytical, 0.1$\times$ solar metallicity & 35, 31 & 33.4 & 47.6\\
~~without Na & 35, 32 & 47.4 & 61.6\\
~~without K & 35, 32 & 57.7 & 71.9\\
\tableline\tableline
\\
\end{tabular}
\tablenotetext{a}{Although the BIC is not a particularly meaningful concept for fixed forward models with one free parameter, we include it here in order to facilitate comparisons between this set of forward models and the analytical models which include either 3 or 4 free parameters.}
\end{center}
\end{table}

\section{Discussion}

WASP-39b is among the coolest gas giant planets with a measured optical transmission spectrum. A visual comparison of P-T profiles for WASP-39b from 1D \citep{sing16} or 3D \citep{kat16} models to those of other gas giant planets shows that its profile is most similar to WASP-6b and HD~189733b at the pressures probed in transmission. However, the measured transmission spectra of these systems are quite different. The transmission spectrum of HD~189733b displays a steep scattering slope across the optical and into the infrared, with evidence for narrow-line abosorption in the Na and K line cores and a muted 1.5~$\upmu$m H$_2$O feature \citep[e.g.][]{hui12, pon13, mcc14}. WASP-6b displays a shallow scattering slope extending to 5~$\upmu$m, with tentative evidence for narrow-line absorption from Na and K in the line cores \citep{nik15}. The transmission spectra of WASP-6b and HD~189733b are best matched by scenarios including high-altitude hazes, while WASP-39b has a shallow optical scattering slope and is well-matched by haze-free models. WASP-39b also appears to possess broad line absorption wings for both Na and K that are absent in the transmission spectra of the other two planets. The differences between these systems are difficult to explain using conventional condensate cloud models, which predict that clouds will form whenever the planet's pressure-temperature profile crosses the condensation curve for a given refractory species \citep{mor13, sing16}.

\begin{table*}
\begin{center}
\caption{Comparison of cooler gas giant systems from the \textit{\scriptsize HST} survey.\label{comparison}}
\begin{tabular}{lccccccl}
\tableline\tableline
\\
System & \Teq~(K) & Mass (\Mj) & $g$~(cm~s$^{-2}$) & Host Star [Fe/H] & \logr \tablenotemark{a} & Clear? & References\tablenotemark{b}\\
\tableline
\\
HD~189733b & 1200 & 1.14 & 2140 & -0.03 $\pm$ 0.08 & -4.501 & no & 1, 2, 3, 4\\
WASP-39b & 1120 & 0.28 & 407 & -0.12 $\pm$ 0.1 & -4.994 & yes & 1, 5; this work\\
WASP-6b & 1150 & 0.50 & 871 & -0.20 $\pm$ 0.09 & -4.741 & no & 1, 6, 7\\
HAT-P-12b & 960 & 0.21 & 562 & -0.29 $\pm$ 0.05 & -5.104 & no & 1, 8, 9, 10\\
\tableline\tableline
\end{tabular}
\tablenotetext{a}{Values are calculated as described in \citet{knu10, isa10} }
\tablenotetext{b}{References: 1. \citet{sing16}, 2. \citet{torres08} 3. \citet{pon13}, 4. \citet{mcc14}, 5. \citet{fae11}, 6. \citet{gil09}, 7. \citet{nik15}, 8. \citet{har09}, 9. \citet{line13}, 10. \citet{mallonn15} }
\end{center}
\end{table*}

It is possible that the different cloud properties of these three planets can be understood through variations in their planet and stellar parameters. We explore possible physical understanding through trends with several model-independent parameters (Table~\ref{comparison}) for these three systems and HAT-P-12b, the only cooler system in the $HST$ survey. For example, higher metallicity would lead to higher concentrations and partial pressures of possible condensate species. Although we do not currently have direct measurements of the atmospheric metallicities of these three planets, we can extrapolate assuming an inverse correlation between mass and metallicity as seen for the solar system gas giants \citep[e.g.][]{wong04,fletcher09,karkoschka11,sromovsky11} and WASP-43b \citep{kre14}. This is supported by current mass and radius measurements for extrasolar planets, which suggest an inverse relationship between mass and bulk metallicity \citep{mil11, tho15}. Surface gravity is also proportional to mass, and lower surface gravity should correspond to lower settling rates and correspondingly longer lifetimes for condensates at high altitudes.  

If metallicity or surface gravity is the dominant factor in the formation of high altitude clouds, we would expect less massive planets to have thicker hazes than their more massive counterparts. WASP-39b is the least massive and therefore the most likely to be metal-rich according to this empirical scaling relation, and yet has the clearest atmosphere in transmission. For these three planets, the relative amount of haze appears to increase with increasing planet mass. However, this trend fails to predict the transmission spectrum observed for HAT-P-12b, which is slightly less massive and cooler than WASP-39b and yet appears to possess a thick cloud or haze layer. Alternatively, we consider the stellar metallicities of the three host stars as a proxy for the likely atmospheric metallicities of the planets. We find no evidence for a correlation between the observed haze opacity and the stellar metallicity, although we note that the metallicities of all three stars are consistent at the 1-2$\sigma$ level.  

The variations in cloud properties could also be attributed to differences in vertical mixing. In this scenario condensate clouds might form at the same pressure in all three atmospheres but could have very different vertical distributions depending on the relative efficiency of vertical mixing versus settling rates in the upper atmosphere. For example, \citet{par13} used a general circulation model to study advection and settling of cloud particles on hot Jupiter HD~209458b and found that particles smaller than a few microns can easily be lofted vertically across many scale heights. The degree of vertical mixing will generally increase as temperature or incident flux increases \citep[e.g.][]{sho10}. If variations in vertical mixing are the dominant factor in controlling the presence of high altitude condensates, we might expect their presence to correlate with temperature. WASP-39b is marginally cooler than HD~189733b and WASP-6b and has a clear atmosphere, roughly consistent with this hypothesis. However this trend does not extend to slightly lower temperatures, as HAT-P-12b is cooler and hosts a thick cloud or haze layer.

Cold traps at depth in the atmosphere might also provide a means to remove condensate cloud particles from the upper atmosphere \citep{spi09}. At these high pressures the vertical mixing rates are likely to be orders of magnitude smaller than in the upper atmosphere, making it difficult to efficiently transport particles upward and depleting the upper atmosphere of the condensate species in question. If this deep cold trap was stronger on WASP-39b than on WASP-6b or HD~189733b, it would provide a natural means to suppress the formation of condensate clouds in WASP-39b's upper atmosphere; this could be explored with planet-specific thermal evolution models \citep[e.g.][]{for08,liu08}.

We also consider that the temperature is expected to vary widely from equator to pole and across terminators \citep[e.g.][]{sho08,sho09,sho15,hen15,kat16}. In this scenario, differences in atmospheric circulation patterns between the three planets might contribute to the presence or absence of localized cloud layers near the day-night terminator \cite[e.g.][]{par13,web15}. Circulation models of these four systems suggest east terminators hundreds of K warmer than western terminators, with the west terminators favorable for the formation of ZnS, KCl, and Na2S clouds \citep{kat16}. Therefore, the differences in observed transmission spectra between WASP-39b and similar systems could plausibly arise from differences in horizontal and vertical mixing, which sets the formation and transport of clouds. Visible-light phase curve observations of hot Jupiters in the Kepler field indicate that a subset of hot Jupiters do indeed possess spatially inhomogenous cloud layers \citep{demo13, shp15}, providing additional support for this hypothesis.

For planets at these relatively low temperatures, photochemical hazes \citep{zah09, mor13} could provide a viable alternative to the standard condensate cloud models. Equally important, hydrocarbons form from methane and therefore should only form in atmospheres cool enough for this molecule to exist in the upper atmosphere. For these photochemically derived hazes, we would expect a positive correlation between the presence of haze and the incident UV flux received from the host star. These three host stars are similar in temperature and spectral type, so any variations in their UV flux is likely due to varying activity levels as measured by their \logr values \citep{knu10}. According to this index HD~189733 is the most active, WASP-6 is moderately active, and WASP-39 is quiet. This trend appears promising at first glance, as the strength of the observed hazes in these three planets increases with increasing stellar activity. However, this trend is again broken at slightly lower temperatures; HAT-P-12b is cool enough to have abundant atmospheric methane and has a hazy atmosphere, yet orbits a relatively quiet star (\Teff~=4500~K, \logr~=-5.104). This discrepancy could potentially be resolved if the clouds observed in HAT-P-12b's atmosphere had a different composition (for instance, condensates rather than a photochemical haze) than those observed in the atmospheres of HD~189733b and WASP-6b.

\section{Conclusions}

In this study we analyze three spectroscopic transits of WASP-39b observed with the STIS instrument on $HST$ (290-1025~nm) and combine these observations with 3.6 and 4.5~$\upmu$m IRAC transit photometry from the $Spitzer~Space~Telescope$. We find that the resulting transmission spectrum is well-matched by models with a Rayleigh scattering slope and Na and K absorption in the optical, in good agreement with the preliminary analysis of these data presented in \citet{sing16}. This is the first system in which the broad wings of both lines are clearly detected. The transmission spectrum is well matched by a clear, H$_2$-dominated atmosphere with either a solar metallicity or sub-solar (0.1$\times$) metallicity, or a solar metallicity model with a weak haze layer. These models predict that this planet should have strong water absorption features in its infrared transmission spectrum; this will be tested by upcoming observations with the WFC3 instrument on $HST$.

WASP-39b is currently one of the coolest gas giant planets with a complete optical transmission spectrum from $HST$ STIS observations. At the altitudes probed in transmission, WASP-39b is very similar in pressure and temperature to HD~189733b and WASP-6b. Interestingly, the latter systems both seem to require a high-altitude haze or cloud layer in order to match their observed transmission spectra. This suggests that the three-dimensional temperature structure must be carefully considered when interpreting and predicting cloud properties. Other factors that may contribute to the presence or absence of high-altitude clouds and hazes include metallicity, surface gravity, vertical mixing, and stellar activity. However, estimates of these parameters based on currently available measurements such as planet mass, temperature, and stellar activity level do not appear to provide a satisfactory explanation for the apparent absence of clouds in WASP-39b's atmosphere as compared to other similar planets. The formation of high-altitude clouds and hazes may also be influenced by the atmospheric circulation patterns and thermal evolution histories of these planets, making them difficult to predict using the relatively simple 1D models described here. The lack of a correlation between the transmission spectra of HD 189733b, WASP-6b, and WASP-39b, as well as the overall diversity of the hot Jupiter transmission spectra obtained to date, suggests that the presence or absence of clouds in these atmospheres most likely results from a combination of multiple parameters and further emphasizes the value of large surveys for developing a better understanding of the processes that drive cloud formation.

\acknowledgments

This work is based on observations with the NASA/ESA $HST$, obtained at the Space Telescope Science Institute (STScI) operated by AURA, Inc. This work is also based in part on observations made with the $Spitzer~Space~Telescope$, which is operated by the Jet Propulsion Laboratory, California Institute of Technology under a contract with NASA. The research leading to these results has received funding from the European Research Council under the European Union's Seventh Framework Program (FP7/2007-2013) / ERC grant agreement no. 336792. Support for this work was provided by NASA through grants under the HST-GO-12473 program from the STScI. G.W.H. and M.H.W. acknowledge long-term support from Tennessee State University and the State of Tennessee through its Centers of Excellence program. We also thank the anonymous referee for suggestions that greatly improved this paper.

\clearpage

\bibliography{mybib}
\end{document}